\documentclass[namedreferences]{solarphysics}
\usepackage[hyperref,optionalrh,showbiblabels]{spr-sola-addons} 
\usepackage{graphicx}                    
\usepackage{amssymb}                     
\usepackage{color}                       
\usepackage{url}                         
\usepackage{breakurl}                    
\usepackage[utf8]{inputenc}
\usepackage[T1]{fontenc}
\usepackage{textcomp}
\usepackage{siunitx}
\usepackage{tabularx,booktabs}
\usepackage{adjustbox}
\usepackage{lipsum}
\usepackage{float}
\usepackage{enumerate}
\usepackage{caption}
\usepackage{threeparttablex}
\usepackage{ulem}
\usepackage{xfrac}
\usepackage{multirow}
\usepackage{rotating}
\usepackage[T1]{fontenc}
\usepackage{romanbar}

\begin{document}

\begin{article}

\begin{opening}

\title{Time Series Analysis of Photospheric Magnetic Parameters of Flare-quiet \textit{versus} Flaring Active Regions: Scaling Properties of Fluctuations}


\author[addressref=aff1,email={eo-jin.lee@khu.ac.kr}]{\inits{E.-J.}\fnm{Eo-Jin}~\lnm{Lee}\orcid{0000-0002-5815-3339}}
\author[addressref=aff2,corref,email={shpark@isee.nagoya-u.ac.jp}]{\inits{S.-H.}\fnm{Sung-Hong}~\lnm{Park}\orcid{0000-0001-9149-6547}}
\author[addressref={aff1,aff3},email={moonyj@khu.ac.kr}]{\inits{Y.-J.}\fnm{Yong-Jae}~\lnm{Moon}\orcid{0000-0001-6216-6944}}


\address[id=aff1]{School of Space Research, Kyung Hee University, Yongin, Republic of Korea}
\address[id=aff2]{Institute for Space-Earth Environmental Research (ISEE), Nagoya University, Nagoya, Japan}
\address[id=aff3]{Department of Astronomy and Space Science, Kyung Hee University, Yongin, Republic of Korea}


\runningauthor{E.-J. Lee \textit{et al.}}
\runningtitle{Time Series Analysis of Active Region Photospheric Magnetic Parameters}


\begin{abstract}
Time series of photospheric magnetic parameters of solar active regions (ARs) are used to answer whether scaling properties of fluctuations embedded in such time series help to distinguish between flare-quiet and flaring ARs. We examine a total of 118 flare-quiet and 118 flaring AR patches (called HARPs), which were observed from 2010 to 2016 by the \textit{Helioseismic and Magnetic Imager} (HMI) on board the \textit{Solar Dynamics Observatory} (SDO). Specifically, the scaling exponent of fluctuations is derived applying the Detrended Fluctuation Analysis (DFA) method to a dataset of 8-day time series of 18 photospheric magnetic parameters at 12-min cadence for all HARPs under investigation. We first find a statistically significant difference in the distribution of the scaling exponent between the flare-quiet and flaring HARPs, in particular for some space-averaged, signed parameters associated with magnetic field line twist, electric current density, and current helicity. The flaring HARPs tend to show higher values of the scaling exponent compared to those of the flare-quiet ones, even though there is considerable overlap between their distributions. In addition, for both the flare-quiet and flaring HARPs the DFA analysis indicates that (1) time series of most of various magnetic parameters under consideration are non-stationary, and (2) time series of the total unsigned magnetic flux and the mean photospheric magnetic free energy density in general present a non-stationary, persistent property, while the total unsigned flux near magnetic polarity inversion lines and parameters related to current density show a non-stationary, anti-persistent trend in their time series.
\end{abstract}


\keywords{Active Regions, Magnetic Fields; Flares, Relation to Magnetic Field}

\end{opening}

\section{Introduction}
\label{s:int}
Solar active regions (ARs) have been intensively studied in the last few decades in the context of how their magnetic field properties are related to flare productivity, mainly using a snapshot of a photospheric line-of-sight or vector magnetogram \citep[\textit{e.g.}][]{2003ApJ...595.1296L,2007ApJ...656.1173L,2007ApJ...661L.109G,2011SpWea...9.4003F,2015ApJ...798..135B,2017ApJ...834...56T}, as well as a sequence of magnetograms over an interval of a few hours to days \citep[\textit{e.g.}][]{2009ApJ...705..821W,2008ApJ...686.1397P,2010ApJ...718...43P,2018SoPh..293..159L}. As a result, it was found that magnetic fields of flaring ARs are more likely to exhibit: (1) morphological complexity (\textit{e.g.} highly fragmented delta-sunspots), (2) higher degrees of magnetic non-potentiality in a wide variety of forms (\textit{e.g.} magnetic shear, electric currents, free magnetic energy), and (3) dynamic evolution in the pre-flare state (\textit{e.g.} rapid flux emergence, sustained cancellation at sheared magnetic polarity inversion lines). Even though such AR magnetic field studies advanced our understanding of flare-productive ARs \citep[see review by][and references therein]{2019LRSP...16....3T}, many efforts are still under way to better discriminate between flare-quiet and flaring ARs.

There have been time series studies of AR photospheric magnetic parameters to find out differences in temporal variations of magnetic field properties between ARs at different levels of flare activity, in particular with respect to flare energy build-up and triggering mechanisms over the course of AR evolution. For example, the superposed epoch analysis \citep[\textit{e.g.}][]{2010ApJ...723..634M,2016ApJ...820..103S} was applied to time series of some photospheric magnetic parameters of flaring ARs, in order to examine any characteristic trend in the time series around the flare occurrence time. Recently, examining 8-day time series of various photospheric magnetic parameters for a total of 93 flare-productive ARs, \citet{2018SoPh..293..159L} found that the examined time series tend to show larger mean values as well as larger fluctuations over a wide range of timescales (\textit{i.e.}, tens of minutes to 1 day) for ARs at higher levels of flare productivity.

Extending the study by \citet{2018SoPh..293..159L}, we examine here the following specific question: ``Do scaling properties of fluctuations presented in time series of AR photospheric magnetic parameters help to distinguish between flare-quiet and flaring ARs?'' Note that scaling properties (also called temporal fractal properties) of time series can be measured by scaling exponents which represent the relationship between fluctuations at different timescales. The scaling exponent increases as fluctuations at shorter/longer timescales become smaller/larger. The scaling exponents can also be considered as a quantitative measure of long-range correlation or persistence in time series: \textit{i.e.}, the higher the scaling exponent, the higher the level of long-range correlation or persistence. Such scaling properties have been studied across a wide variety of different time series \citep[\textit{e.g.}][]{1997PhyA..246..454V,Eke2000,2005PhRvL..94a0602L,2009SoPh..260..441S,2013MNRAS.436.2907M}, in particular to identify an underlying physical mechanism which may be responsible for the scaling property estimated from a given time series \citep[\textit{e.g.}][]{2010PhRvE..81c1101M}. In this respect, we speculate that scaling properties of time varying AR magnetic properties may capture a difference in the magnetic field evolution of flare-quiet \textit{versus} flaring ARs if any exists. For example, in the case of temporal variations in the total unsigned magnetic flux, flaring ARs may exhibit higher values of scaling exponents compared to those of flare-quiet ARs, because flaring ARs often show persistent flux variations (either increase or decrease in relation to flux emergence or cancellation) over timescales of a few hours to days before flare occurrence.

To address the targeted question of this study, we examine 8-day time series of numerous photospheric magnetic parameters for a total of 236 different AR patches (\textit{i.e.} 118 flare-quiet and 118 flaring AR patches). Specifically, scaling exponents of fluctuations in the AR photospheric magnetic parameter time series are estimated using the Detrended Fluctuation Analysis \citep[DFA;][]{1994PhRvE..49.1685P,1995Chaos...5...82P}. DFA is a widely used method to quantify scaling exponents of various time series datasets \citep[\textit{e.g.}][]{1994PhRvE..49.1685P,1995Chaos...5...82P,1997PhyA..246..454V,2002PhRvE..65e1908K,2005MAP....88..119K,TELESCA2008151,10.3389/fphys.2012.00450,doi:10.1098/rsif.2013.1112,2015RemS....712942G}. Details of how DFA is applied to the magnetic parameter time series are explained in Section~\ref{s:data}. The analysis results are shown in Section~\ref{s:results}, including (1) distribution of DFA scaling exponents for each of the examined photospheric magnetic parameter time series and (2) comparison of DFA scaling exponents between the flare-quiet and flaring AR patches under study. In Section~\ref{s:sum}, we discuss why scaling exponents of different magnetic parameter time series show different levels of discriminating the flaring AR patches from the others.

\section{Data and Analysis}
\label{s:data}
The \textit{Helioseismic and Magnetic Imager} \citep[HMI;][]{2012SoPh..275..207S} on board the \textit{Solar Dynamics Observatory} \citep[SDO;][]{2012SoPh..275....3P} routinely produces full-disk, photospheric vector magnetograms with a spatial resolution of 0.5 arcsec per pixel and 12-min cadence. A total of 18 photospheric magnetic parameters of ARs are derived from HMI photospheric vector magnetic field data and available as part of the Space-weather HMI Active Region Patch/Cylindrical Equal Area \citep[SHARP/CEA;][]{2014SoPh..289.3549B} products. Details of the 18 SHARP magnetic parameters can be found in Table~\ref{table_1}. In this study, we use the 18 SHARP parameters and their time series. Note that the SHARP parameters are determined for a given HMI AR Patch \citep[HARP;][]{2014SoPh..289.3483H}, only considering high-confidence pixels (\textit{i.e.} HMI FITS header keyword {\tt CONF\_DISAMBIG}\,$=$\,90) within a smooth bounding curve (\textit{i.e.} HMI keyword {\tt BITMAP}\,$\ge$\,33) defined inside the rectangular HARP bounding box. The total number of high-confidence pixels in HARPs tends to significantly decrease as they are located closer to the solar limb \citep{2014SoPh..289.3483H,2015ApJ...798..135B}. We therefore restricted here the use of the SHARP parameters, as calculated only from HARPs of which the centre positions are within $\pm$60$^\circ$ from the central meridian (hereafter, referred to as $\Theta_{60}$).

We searched for HARPs that were continuously observed over 8 days within $\Theta_{60}$ during an interval from May 2010 to February 2016. We first found a set of 118 flaring HARPs, in which each HARP produced at least one M- or X-class flare throughout its passage of $\Theta_{60}$. For comparison, the other set of 118 flare-quiet HARPs that never produced any flares at and above C-class over their entire solar disk passage was also collected; it was constructed in the way that the total number of the flare-quiet HARPs per year is almost equally distributed over the years of the same interval under consideration to search for the flaring HARPs. Each of the flare-quiet HARPs under investigation is assigned with one NOAA AR; however, in the case of the flaring HARPs, half of them contain two or more NOAA ARs for a given HARP. A total of the 236 HARPs consisting of the 118 flare-quiet and 118 flaring HARPs were selected in this study to examine the targeted question of whether there are any differences in scaling properties of SHARP parameter time series of flare-quiet \textit{versus} flaring ARs.

\begin{figure}
\centering
\includegraphics[width=\textwidth]{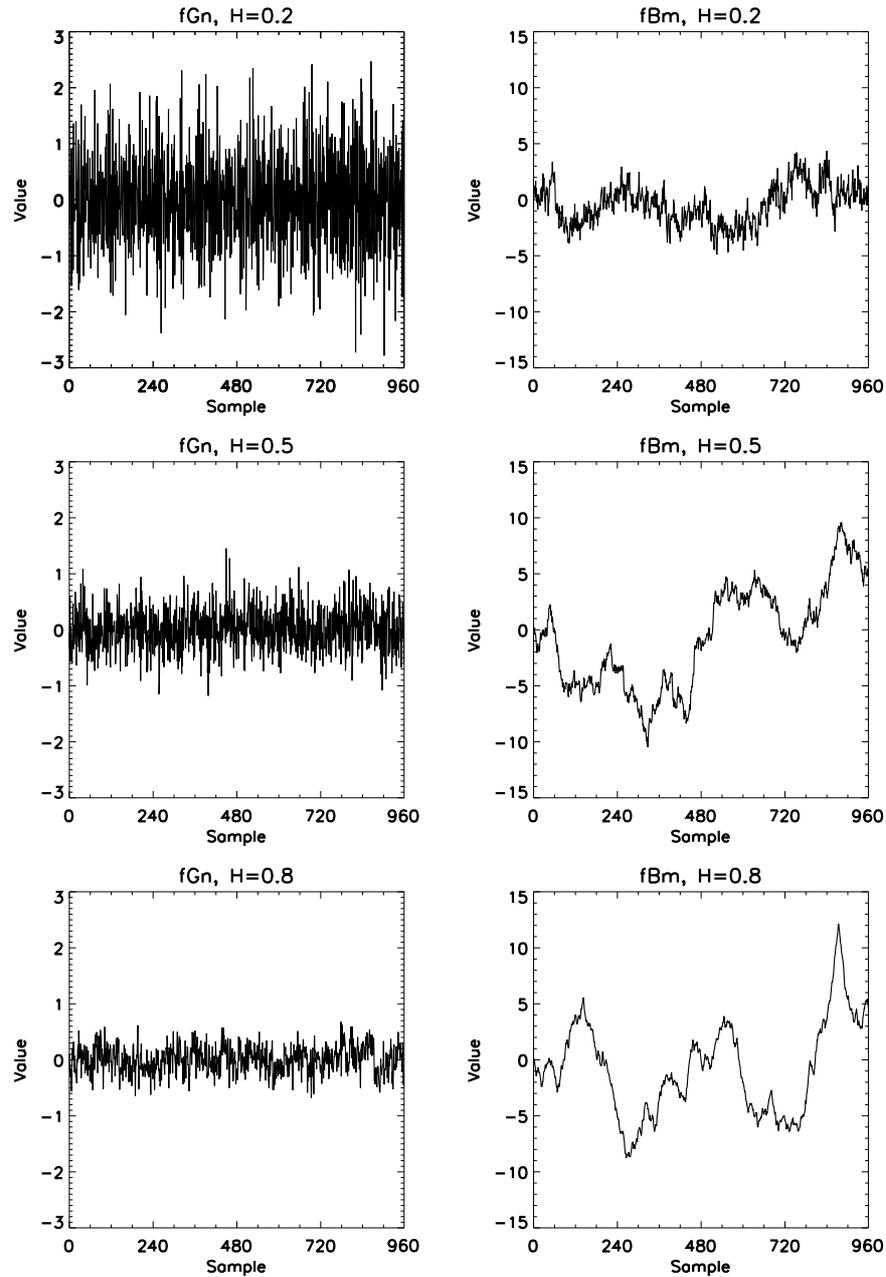}
\caption{Examples of the fractional Gaussian noise (fGn) and fractional Brownian motion (fBm) for three different values of Hurst scaling exponent $H$: (\textit{top}) $H$\,$=$\,0.2, (\textit{middle}) $H$\,$=$\,0.5, and (\textit{bottom}) $H$\,$=$\,0.8. The fBm time series (\textit{right column}) are synthesized by the ``wfbm'' function in the Matlab and the fGn time series (\textit{left column}) are calculated from successive increments of the fBm time series.}
\label{f1}
\end{figure}

For each of the 236 HARPs under consideration, we analyzed a set of 8-day time series of the 18 SHARP parameters using the same analysis method as in \citet{2018SoPh..293..159L}. Each 8-day time series $\textbf{X}$ of the SHARP parameters consists of 960 data points at 12-min cadence: \textit{i.e.} $\textbf{X}$\,$=$\,$\left[\mathrm{X}_{1},\mathrm{X}_{2},\mathrm{X}_{3},\ldots, \mathrm{X}_{960}\right]$. For the time series $\textbf{X}$, the scaling exponent is calculated using the DFA method. A major advantage of the DFA method is that it is designed to determine the scaling exponent of any time series regardless of whether the underlying statistics (such as mean and variance) of the time series present stationary (constant over time) or non-stationary (changing with time) characteristics. \citet{1968SIAMR..10..422M} introduced the fractional Gaussian noise (fGn) and fractional Brownian motion (fBm) to model stationary and non-stationary time series, respectively, with the so-called Hurst scaling exponent $H$. The Hurst exponent is a measure of long-term memory (or persistence) of a given time series, and it ranges from 0 to 1. In principle, successive increments of a non-stationary fBm time series with a given value of $H$ can be considered as a stationary fGn time series with the same $H$ value \citep{2016PhyA..461..662T}.
Time series can be in general classified into three different categories, depending on the values of the Hurst exponent derived from the given time series: anti-persistent (0\,$<$\,$H$\,$<$\,0.5), random ($H$\,$=$\,0.5), and persistent (0.5\,$<$\,$H$\,$<$\,1). Figure~\ref{f1} shows examples of both fGn (left panels) and fBm (right panels) times series with different Hurst exponents. In general, time series of both fGn and fBm have a tendency to be more rough as $H$ gets closer to 0, while more smooth as $H$ is closer to 1 \citep{10.3389/fphys.2012.00450}. As shown in Figure~\ref{f1}, anti-persistent time series (top panels) show a tendency of an increase followed by a decrease (or vice versa) over a short timescale (\textit{i.e.} large fluctuation). On the other hand, persistent time series (bottom panels) tend to exhibit an increase followed by a decrease (or vice versa) over a longer timescale. Meanwhile, successive variations on random time series (middle panels) are uncorrelated to each other. The scaling exponent $\alpha$ calculated from the DFA method has been related to the Hurst exponent \citep[\textit{e.g.}][]{Eke2000,10.3389/fphys.2012.00450}: $H$\,$=$\,$\alpha$ in the case of fGn time series and $H$\,$=$\,$\alpha$\,$-$\,1 for fBm time series. In the present study, the scaling property of a given SHARP parameter time series refers to the DFA scaling exponent $\alpha$.

\begin{figure}
\centering
\includegraphics[width=\textwidth]{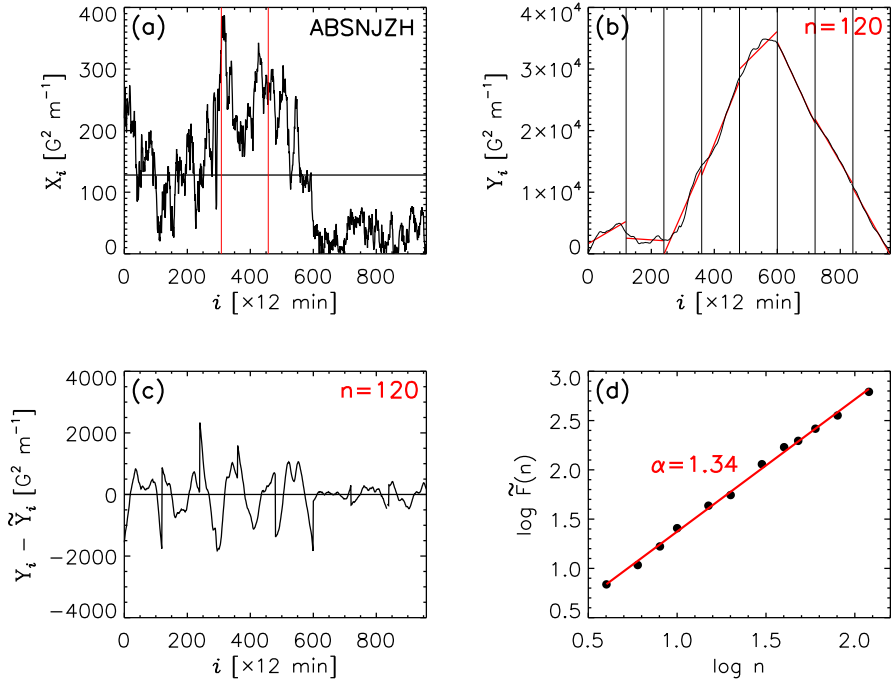}
\caption{Example of calculating the DFA exponent $\alpha$. \textit{Panel} (\textbf{a}) shows the 8-day time series $\textbf{X}$ (\textit{black line}) of the SHARP parameter ABSNJZH (\textit{i.e.} the absolute value of the net vertical current helicity) for the HARP number 5011. The mean of $\textbf{X}$ is shown by the \textit{horizontal line}, and the start times of two major flares (M1.0 above) produced in this HARP by the \textit{vertical red lines}. \textit{Panel} (\textbf{b}) shows the cumulative sum $\textbf{Y}$ (\textit{black line}) of $\textbf{X}$, and a set of local linear trends $\mathrm{\tilde{Y}}$ (\textit{red lines}) for non-overlapping segments of equal length $n$\,$=$\,120. The deviation of $\textbf{Y}$ from $\mathrm{\tilde{Y}}$ is shown in \textit{Panel} (\textbf{c}). \textit{Panel} (\textbf{d}) shows a log-log plot of the detrended fluctuation $\mathrm{\tilde{F}}$ (\textit{i.e.} the root-mean-square of $(\textbf{Y}-\mathrm{\tilde{Y}})$) as a function of $n$. The DFA exponent $\alpha$ is defined as the slope of the least-squares straight-line fit (\textit{red line}) of the log-log plot. \textit{Panels} (\textbf{a}--\textbf{c}) adapted from \citet{2018SoPh..293..159L}.}
\label{f2}
\end{figure}

We now demonstrate in Figure~\ref{f2} how $\alpha$ is calculated from a SHARP parameter time series. Figure~\ref{f2}{a} shows the 8-day time series $\textbf{X}$ of the absolute value of the net current helicity (\textit{i.e.} ABSNJZH in Table~\ref{table_1}) in the HARP number 5011. The cumulative sum $\textbf{Y}$\,$=$\,$\left[\mathrm{Y}_{1},\mathrm{Y}_{2},\mathrm{Y}_{3},\ldots, \mathrm{Y}_{960}\right]$ is first calculated from $\textbf{X}$ by integration of a sequence  $\textbf{X}-\mathrm{\bar{X}}$ as
\begin{equation}
\mathrm{Y}_{k} = \sum_{i=1}^k \left( \mathrm{X}_{i} - \mathrm{\bar{X}} \right),
\end{equation}
where $\mathrm{\bar{X}}$ is the mean of $\textbf{X}$. In Figure~\ref{f2}{b}, $\textbf{Y}$ is shown by the black solid line. $\textbf{Y}$ is then divided into non-overlapping segments of the equal length $n$\,$=$\,120. For each segment, a local linear trend $\mathrm{\tilde{Y}}(n$\,$=$\,120$)$ (red lines in Figure~\ref{f2}{b}) is determined by least-squares straight-line fit. Next the detrended fluctuation $\mathrm{\tilde{F}}(n$\,$=$\,120$)$ is calculated by the root-mean square of the detrended cumulative sum (\textit{i.e.} $\textbf{Y}$\,$-$\,$\mathrm{\tilde{Y}}(n$\,$=$\,120$)$, Figure~\ref{f2}{c}) as
\begin{equation}
\mathrm{\tilde{F}}(n\!=\!120) = \sqrt{\frac{1}{N} \sum_{i=1}^{960} \left[ \mathrm{Y}_{i} - \mathrm{\tilde{Y}}_{i}(n\!=\!120) \right]^2}.
\end{equation}
Calculating $\mathrm{\tilde{F}}(n)$ for a set of $n$\,$=$\,[4, 6, 8, 10, 15, 20, 30, 40, 48, 60, 80, 120], we estimated $\alpha$ as the slope of least-squares straight-line fit between $\mathrm{\tilde{F}}(n)$ and $n$ in a log-log plot (Figure~\ref{f2}{d}).

\begin{figure}
\centering
\includegraphics[width=\textwidth]{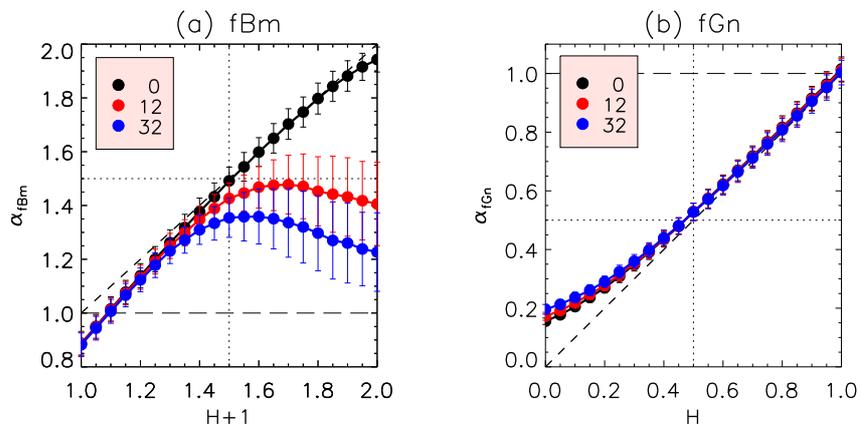}
\caption{Test of estimating $\alpha$ values from a set of (\textbf{a}) fBm and (\textbf{b}) fGn time series that have different values (\textit{i.e.} 0--1 at intervals of 0.05) of $H$ and three different numbers (\textit{i.e.} 0, 12, and 32) of missing data points. For a given set of the $H$ value and the number of missing data points, the mean (\textit{filled circle}) and standard deviation (\textit{error bar}) of estimated $\alpha$ values are calculated from a set of 1,000 fBm and fGn time series, respectively.}
\label{f3}
\end{figure}

Prior to applying the DFA method to SHARP parameter time series, we examined how well the DFA method estimates $\alpha$ for both stationary (fGn) and non-stationary (fBm) time series with a different number of missing data points. For this, fBm time series are synthesized for $H$ values from 0 to 1 at an interval of 0.05, using the ``wfbm'' function in the Matlab \citep{ABRY1996377,bardet:hal-00127929}. For a given $H$ value, we make 1,000 different fBm time series of which each consists of 960 data points. Some data points in the time series are randomly selected and replaced with NaN (Not a Number) values to make missing data points. The number of missing data points in SHARP parameter time series varies depending on the HARP and SHARP parameter under investigation. For all examined HARPs in this study, the number of missing data points included in each USFLUX time series is $\sim$32 on average, while the median is $\sim$12. Figure~\ref{f3}{a} shows the mean and the standard deviation of the DFA scaling exponent $\alpha_\mathrm{fBm}$ estimated from the fBm time series as a function of $H$ for the three different sets of fBm time series with no (black), 12 (red), and 32 (blue) missing data points. For the fBm time series without any missing data point, in general, the linear relationship (black line in Figure~\ref{f3}{a}) between $H$\,$+$\,1 and $\alpha_\mathrm{fBm}$ appears to be strong in a substantial range of $H$\,$+$\,1 values, except for a slight deviation from the dashed diagonal line (indicating a theoretical linear relationship between $H$\,$+$\,1 and $\alpha_\mathrm{fBm}$) in the range of $H$\,$+$\,1\,$<$\,1.3. For the fBm times series with the missing data points, however, this linear relationship does not become valid in the range of $H$\,$+$\,1\,$>$\,1.5. Next, fGn time series are derived from successive increments of the fBm time series. For fGn time series, as shown in Figure~\ref{f3}{b}, a much stronger linear relationship exists between $H$ and $\alpha_\mathrm{fGn}$ for all three different set of fGn time series regardless of the number of missing data points in the time series. Note that $\alpha_\mathrm{fGn}$ is somewhat overestimated (0.05--0.2) compared to the dashed diagonal line in the range of $H$\,$<$\,0.3.

In principle, $\alpha$ from the fBm or fGn time series should have a linear relationship with $H$. However, a deviation appears mainly due to missing data existed in time series and/or uncertainties in estimating $\alpha$ using the DFA method. For the fBm and fGn time series with 12 NaNs, Figure~\ref{f3} shows that the deviation from the dashed diagonal line for $\alpha_\mathrm{fGn}$ becomes larger than that for $\alpha_\mathrm{fBm}$ in the range of $H$\,$<$\,0.32; \textit{i.e.} for $H$\,$=$\,0.32, its corresponding $\alpha_\mathrm{fGn}$ with 12 NaNs is 0.369. We therefore determine $\alpha$ in this study as follows. First, $\alpha_\mathrm{fBm}$ is calculated from a given original SHARP parameter time series. If $\alpha_\mathrm{fBm}$\,$\leq$\,1, then $\alpha_\mathrm{fBm}$ is considered as $\alpha$. On the other hand, in the case of $\alpha_\mathrm{fBm}$\,$>$\,1, $\alpha_\mathrm{fGn}$ is calculated from successive increments of the SHARP parameter time series. If $\alpha_\mathrm{fGn}$\,$<$\,0.369, then $\alpha_\mathrm{fBm}$ is taken into account as $\alpha$; otherwise, $\alpha_\mathrm{fGn}$\,$+$\,1 as $\alpha$.

\section{Results}
\label{s:results}
\begin{figure}
\centering
\includegraphics[width=\textwidth]{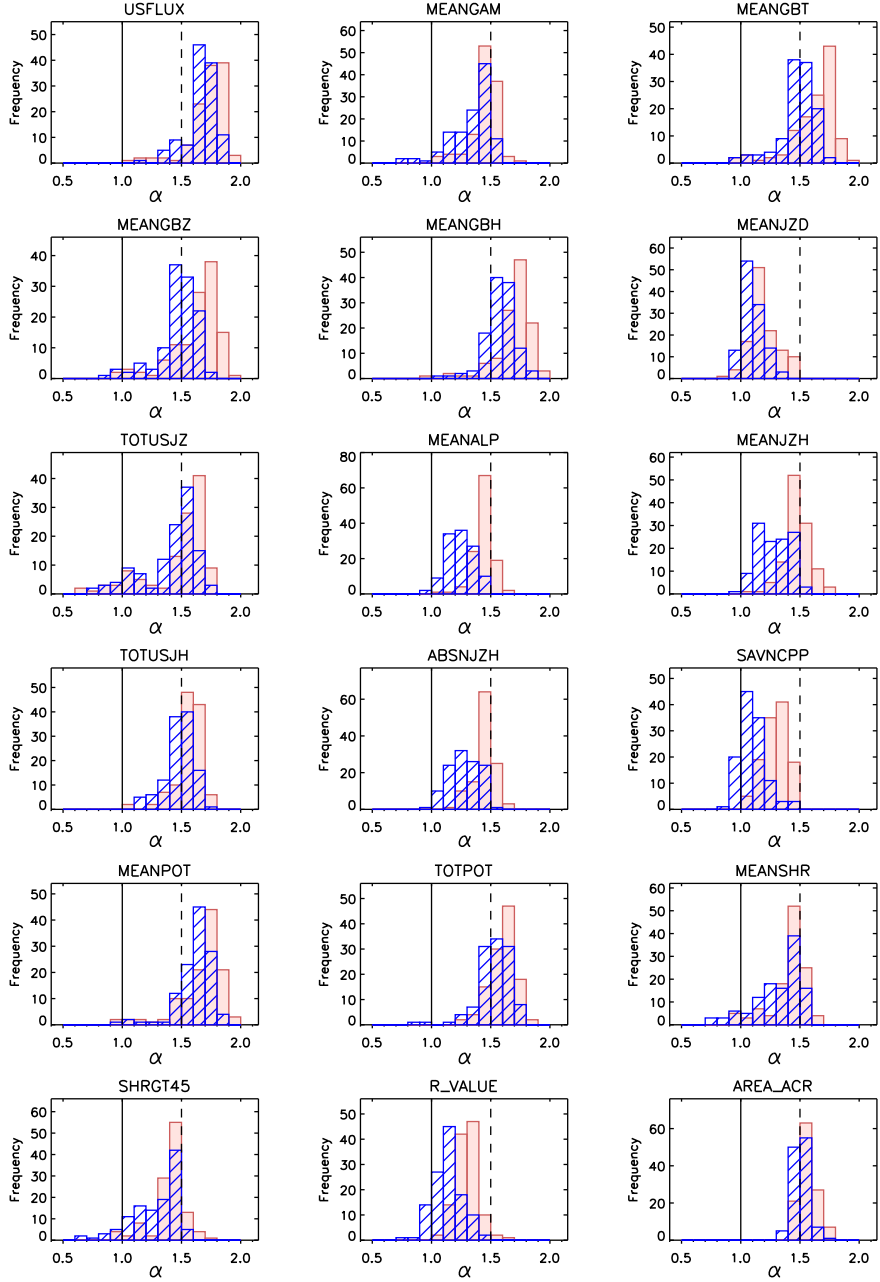}
\caption{Distributions of $\alpha$ values estimated from time series of all 18 SHARP parameters for 118 flare-quiet (\textit{blue bars}) and 118 flaring HARPs (\textit{red bars}), respectively. In \textit{each panel}, the boundary between stationary and non-stationary time series (\textit{i.e.} $\alpha$\,$=$\,1.0) is marked by the \textit{vertical solid line} and the boundary between anti-persistent and persistent property (\textit{i.e.} $\alpha$\,$=$\,1.5) by the \textit{vertical dashed line}.}
\label{f4}
\end{figure}

The DFA scaling exponent $\alpha$ was calculated from each 8-day time series of all 18 SHARP parameters for each of the 118 flare-quiet and 118 flaring HARPs observed from May 2010 to February 2016. For a given SHARP parameter, Figure~\ref{f4} shows the frequency distribution of $\alpha$ values for the flare-quiet (blue bars) and flaring HARPs (red bars), respectively. The distributions of $\alpha$ values tend to be left-skewed for most of the SHARP parameters but right-skewed for the mean vertical current density MEANJZD; on the other hand, in the case of the SHARP parameter R\_VALUE, $\alpha$ values show a symmetric distribution. We define six subgroups according to $\alpha$ values as follows: 0.0\,$\le$\,$\alpha$\,$<$\,0.4 (stationary and anti-persistent), 0.4\,$\le$\,$\alpha$\,$<$\,0.6 (stationary and random), 0.6\,$\le$\,$\alpha$\,$<$\,1.0 (stationary and persistent), 1.0\,$\le$\,$\alpha$\,$<$\,1.4 (non-stationary and anti-persistent, $A$), 1.4\,$\le$\,$\alpha$\,$<$\,1.6 (non-stationary and random, $R$), and 1.6\,$\le$\,$\alpha$\,$<$\,2.0 (non-stationary and persistent, $P$). Table~\ref{table_2} shows percentage distributions of $\alpha$ values categorized into the subgroups for the flare-quiet and flaring HARPs, respectively. Note that none of the flare-quiet and flaring HARPs have $\alpha$ values less than 0.6. The SHARP parameters are grouped as $PP$, $RP$, $RR$, $AR$, or $AA$, considering the subgroup with the highest percentage (marked in bold in Table~\ref{table_2}) for the flare-quiet (first letter) and flaring HARPs (second letter), respectively.

We find that the SHARP parameter time series are predominantly non-stationary for both the flare-quiet and flaring HARPs (refer to Figure~\ref{f4} and Table~\ref{table_2}). This indicates that most of the SHARP parameter time series show some long-term (a few tens of hours to a few days) variations during the disk passage of the HARPs. It is also found that the flaring HARPs tend to show relatively higher $\alpha$ values than those of the flare-quiet HARPs. Moreover, for some SHARP parameter time series, we find the following three characteristics in the context of their persistent, random or anti-persistent properties: (1) the total unsigned magnetic flux USFLUX and the mean photospheric magnetic free energy density MEANPOT show the persistent property in their time series for a majority of the flare-quiet and flaring HARPs, (2) the time series of the mean characteristic twist parameter MEANALP and the two current helicity parameters ABSNJZH and MEANJZH tend to be anti-persistent for the flare-quiet HARPs, while random for the flaring HARPs, and (3) MEANJZD, R\_VALUE, and the sum of the absolute value of the net current per polarity SAVNCPP show a strong tendency of the anti-persistent property for most of the flare-quiet and flaring HARPs.

\begin{figure}
\centering
\includegraphics[width=\textwidth]{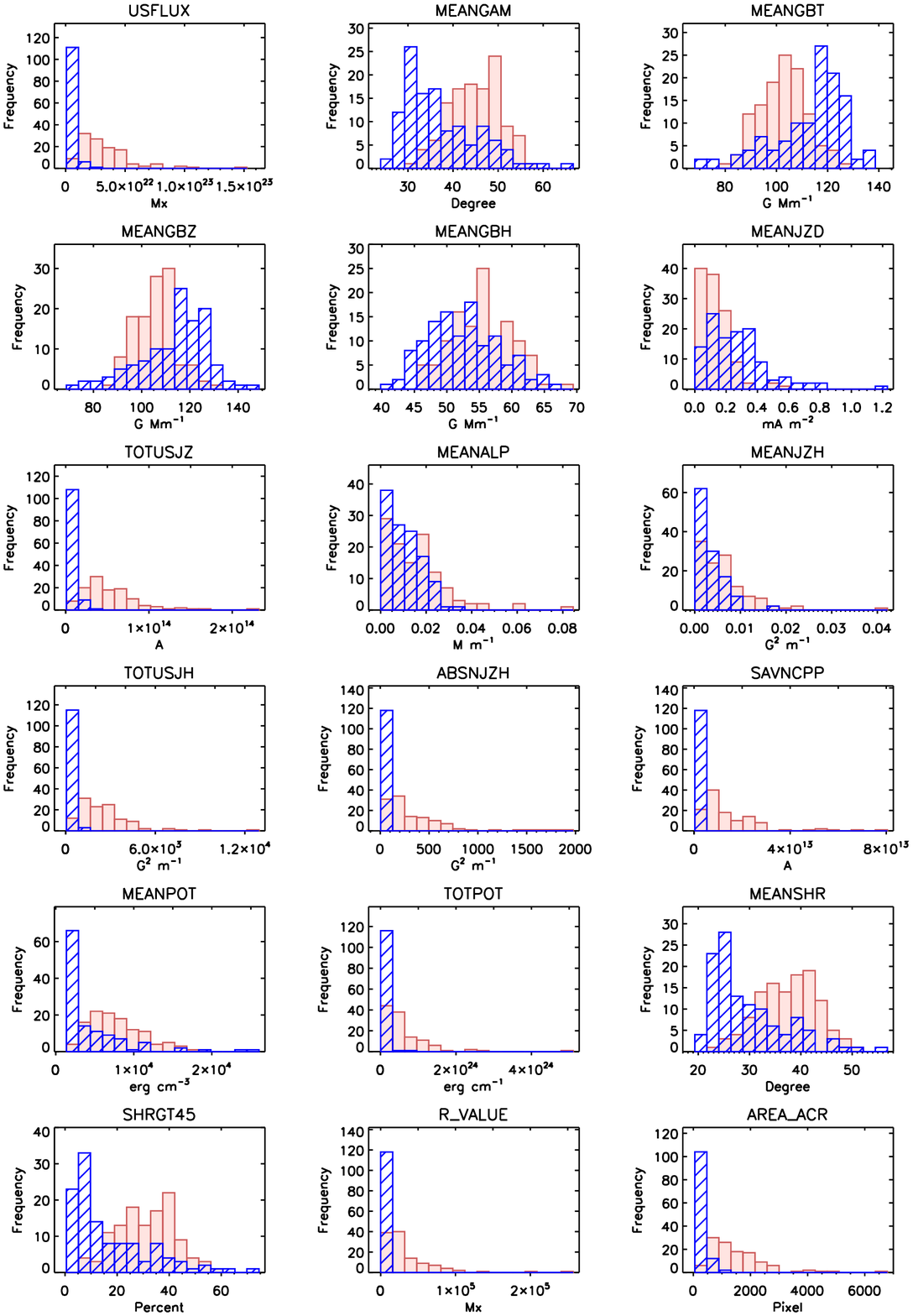}
\caption{Same as Figure~\ref{f4}, but for the mean values calculated from time series of all 18 SHARP parameters for 118 flare-quiet (\textit{blue bars}) and 118 flaring HARPs (\textit{red bars}), respectively.}
\label{f5}
\end{figure}

For comparison with $\alpha$, we take into account the mean $\mathrm{\bar{X}}$ of a given SHARP parameter time series. Figure~\ref{f5} shows in each panel the frequency distribution of $\mathrm{\bar{X}}$ values for the flare-quiet (blue bars) and flaring HARPs (red bars), respectively, for each of the examined SHARP parameters. As shown in the distributions of $\alpha$ values in Figure~\ref{f4}, the flaring HARPs in general show larger $\mathrm{\bar{X}}$ values compared to the flare-quiet HARPs for most of the SHARP parameters, while the opposite trend appears in the case of the mean gradients of the total/vertical fields MEANGBT/MEANGBZ and the mean vertical current density MEANJZD. $\mathrm{\bar{X}}$ values show a right-skewed distribution for many of the SHARP parameters; in many cases, for the flare-quiet HARPs the distributions have a peak at the bin including the minimum value. In addition, we find in many qualitative aspects that the distributions of $\mathrm{\bar{X}}$ values for the flaring HARPs are more differentiated from those for the flare-quiet HARPs with respect to several space-integrated unsigned SHARP parameters such as USFLUX, TOTUSJZ, TOTUSJH, TOTPOT, R\_VALUE, and AREA\_ACR.

\begin{figure}
\centering
\includegraphics[width=\textwidth]{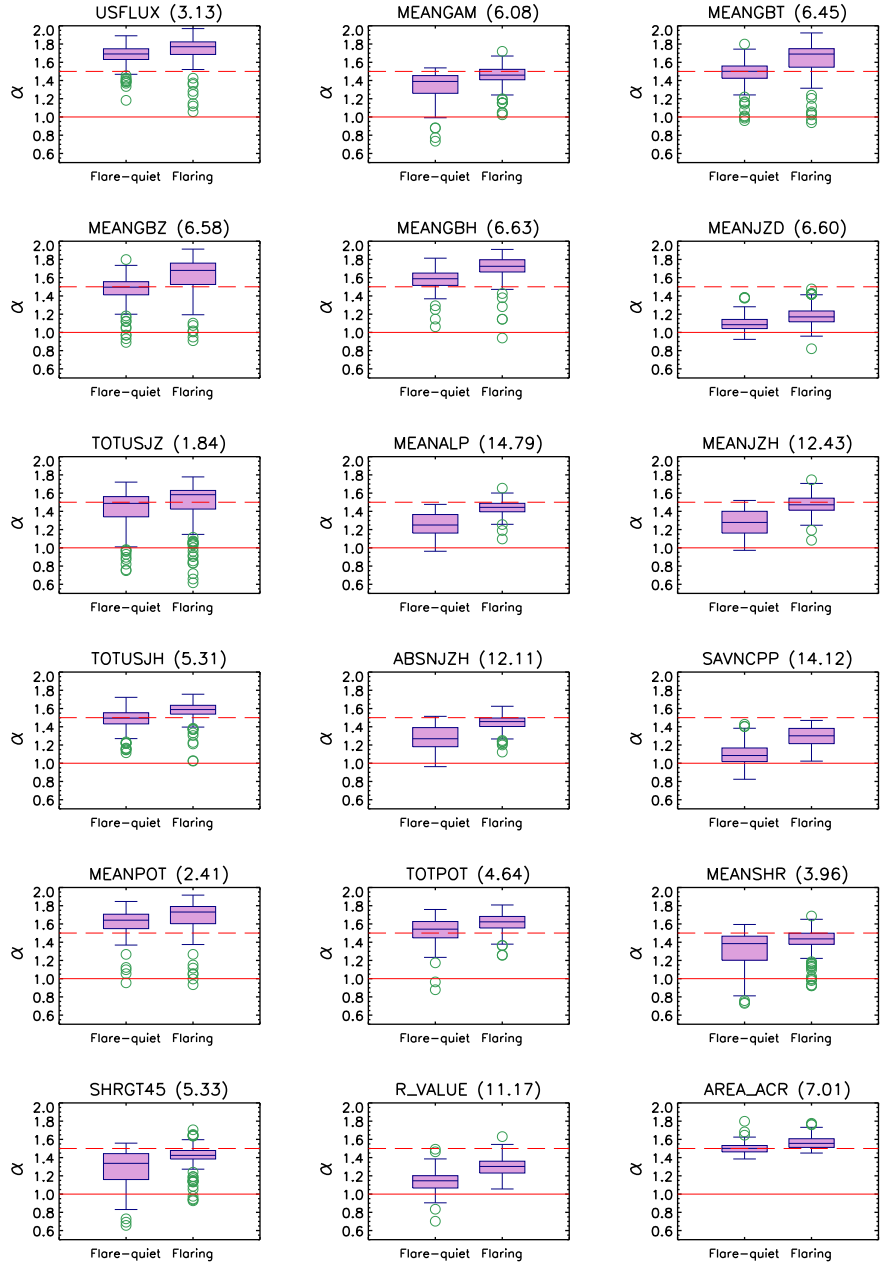}
\caption{Box and whisker plots are used to represent distributions of $\alpha$ values from time series of all 18 SHARP parameters for 118 flare-quiet and 118 flaring HARPs, respectively. The Student's $t$-test result (\textit{i.e.} $t$-value) is shown in parenthesis next to the given SHARP parameter keyword in \textit{each panel}. In \textit{each panel}, the boundary between stationary and non-stationary time series (\textit{i.e.} $\alpha$\,$=$\,1.0) is marked by the \textit{horizontal red solid line} and the boundary between anti-persistent and persistent property (\textit{i.e.} $\alpha$\,$=$\,1.5) by the \textit{horizontal red dashed line}.}
\label{f6}
\end{figure}

\begin{figure}
\centering
\includegraphics[width=\textwidth]{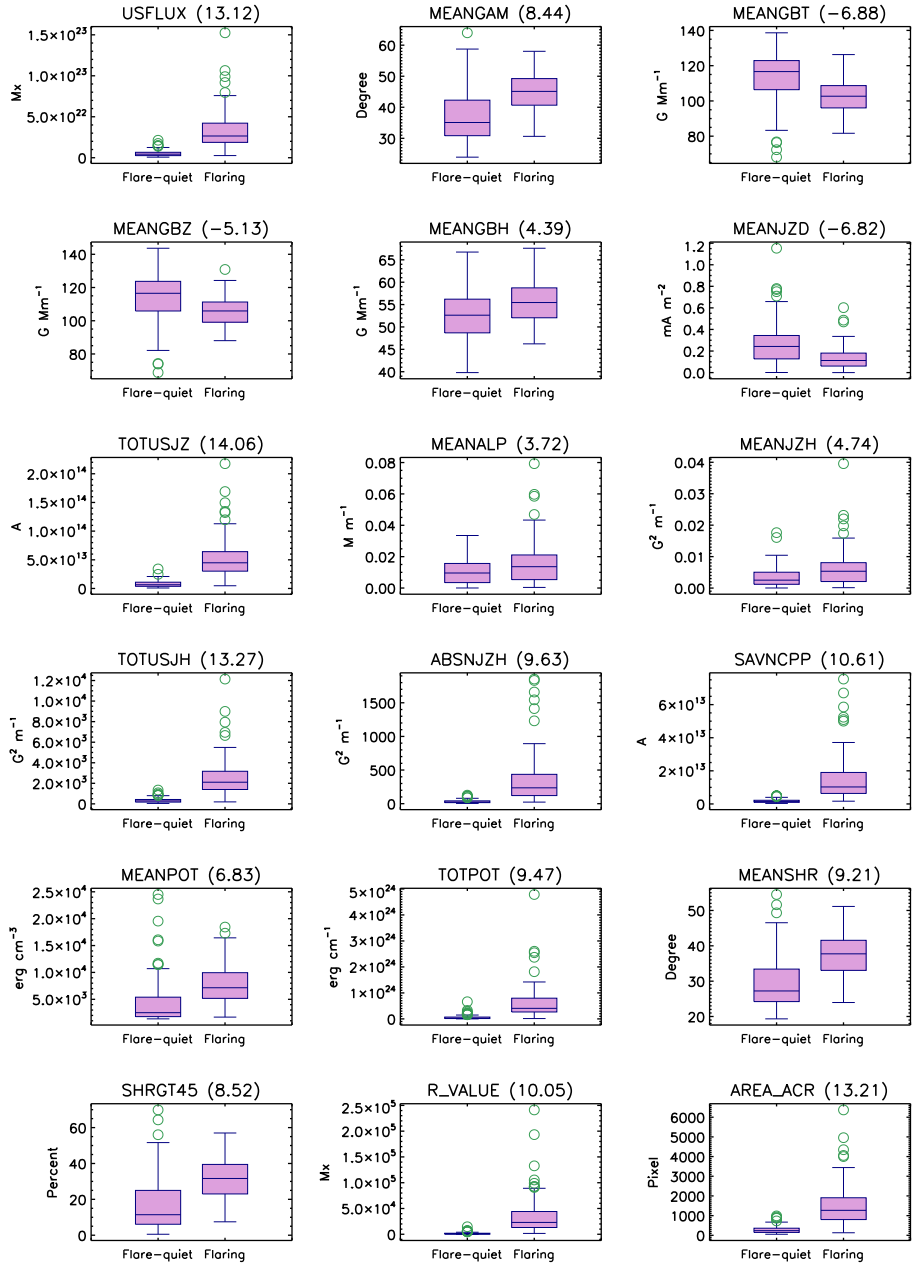}
\caption{Same as Figure~\ref{f6}, but for the mean values calculated from time series of all 18 SHARP parameters for 118 flare-quiet and 118 flaring HARPs, respectively.}
\label{f7}
\end{figure}

Another graphical way to present distribution characteristics of $\alpha$ values is to use a box and whisker plot. In Figure~\ref{f6}, each panel shows box and whisker plots for $\alpha$ values of the flare-quiet (left) and flaring HARPs (right), respectively, calculated from time series of a given SHARP parameter. The horizontal bar inside each box indicates the median of the $\alpha$ distribution, and the bottom and top lines of the box are the first and third quartiles (\textit{i.e.} $\alpha_{\mathrm{1st}}$ and $\alpha_{\mathrm{3rd}}$). If the minimum/maximum value exists within the range extended from the first/third quartile to 1.5\,$\times$\,$[\alpha_{\mathrm{3rd}}$\,$-$\,$\alpha_{\mathrm{1st}}]$ above/below, then the bottom/top whisker outside the box is located at the minimum/maximum value; otherwise, the whiskers are plotted at $\alpha_{\mathrm{1st}}$\,$-$\,1.5\,$\times$\,$[\alpha_{\mathrm{3rd}}$\,$-$\,$\alpha_{\mathrm{1st}}]$ and $\alpha_{\mathrm{3rd}}$\,$+$\,1.5\,$\times$\,$[\alpha_{\mathrm{3rd}}$\,$-$\,$\alpha_{\mathrm{1st}}]$. All data points located outside of the whiskers are marked with circles. Figure~\ref{f6} shows that $\alpha_{\mathrm{1st}}$ for the flaring HARPs is even larger than $\alpha_{\mathrm{3rd}}$ for the flare-quiet HARPs in the case of several SHARP parameters such as MEANGBH, MEANALP, MEANJZH, ABSNJZH, SAVNCPP, and R\_VALUE. In Figure~\ref{f7}, the same box and whisker plots are shown for $\mathrm{\bar{X}}$ values, which can be compared with those for $\alpha$ values in Figure~\ref{f6}. As in Figure~\ref{f6}, we find that the first quartile of $\mathrm{\bar{X}}$ values for the flaring HARPs is larger than the third quartile of those for the flare-quiet HARPs for some SHARP parameters such as USFLUX, TOTUSJZ, TOTUSJH, ABSNJZH, SAVNCPP, TOTPOT, R\_VALUE, and AREA\_ACR. ABSNJZH, SAVNCPP, and R\_VALUE seem to perform well to differentiate between the flare-quiet and flaring HARPs, using $\alpha$ as well as $\mathrm{\bar{X}}$, in the respect of the comparison of the third quartile of the flare-quiet HARPs with the first quartile of the flaring ones. On the other hand, in the context of the potential of discriminating flare-quiet HARPs from flaring ones, a tendency is found that some space-averaged, signed SHARP parameters (\textit{e.g.} MEANALP and MEANJZH) work better in the case of $\alpha$, while space-integrated, unsigned parameters for $\mathrm{\bar{X}}$ (\textit{e.g.} USFLUX, TOTUSJZ, TOTUSJH, TOTPOT, and AREA\_ACR).

We performed the Student's $t$-test to quantitatively examine whether two sets of $\alpha$ values (one for the flare-quiet HARPs and the other for the flaring HARPs), as well as two sets of $\mathrm{\bar{X}}$ values (again, for the flare-quiet and flaring HARPs, respectively), are significantly different from each other. As a result of the $t$-test on the given two datasets, a $t$-value is calculated, which measures the difference between the means of the two datasets divided by the standard error of the mean difference. The calculated $t$-value is presented in parenthesis next to the given SHARP parameter keyword in each panel of Figures~\ref{f6} and~\ref{f7}, respectively. In this study, the sign of the $t$-value is defined to be positive when the mean of the flaring HARPs is larger than that of the flare-quiet HARPs. Throughout the $t$-test, we also calculated the $p$-value, \textit{i.e.}, the probability of accepting the null hypothesis that the means of the two datasets are equal. We first find that $p$-values for all SHARP parameters are less than $\sim$0.05, which indicates there is a statistically significant difference in the mean of the two datasets between the flare-quiet and flaring HARPs for both cases of $\alpha$ and $\mathrm{\bar{X}}$. For most SHARP parameters, the flaring HARPs typically show higher $\alpha$ and/or $\mathrm{\bar{X}}$ values compared to the flare-quiet HARPs so that their $t$-values are positive and in the range of 2 to 14. Table~\ref{table_3} shows the $t$-test results (\textit{i.e.}, $t$-values and $p$-values) for all SHARP parameters, and four best-discriminating parameters are highlighted for $\alpha$ and $\mathrm{\bar{X}}$, respectively. Interestingly, as found in the box and whisker plots, the same trend appears that the space-averaged, signed parameters, MEANALP and MEANJZH, are listed as the best-discriminating parameters in the case of $\alpha$, while for $\mathrm{\bar{X}}$ the space-integrated, unsigned parameters, TOTUSJZ, TOTUSJH, AREA\_ACR, and USFLUX.

\section{Summary and Discussions}
\label{s:sum}
This study was motivated from the question of whether scaling properties of fluctuations in AR photospheric magnetic parameter time series help to distinguish between flare-quiet and flaring ARs. To answer the question, using the DFA method, we determined the scaling exponent $\alpha$ of each 8-day time series of all 18 SHARP parameters for each of 118 flare-quiet and 118 flaring HARPs observed from 2010 to 2016. The mean $\mathrm{\bar{X}}$ calculated from the same 8-day time series was also considered for comparison. Our major findings can be summarized as follows:
\begin{enumerate}[(i)]
\item{The distributions of the DFA scaling exponent $\alpha$ between the flare-quiet and flaring HARPs show a statistically significant difference, even if there is considerable overlap between their distributions, in particular in the case of the space-averaged, signed SHARP parameters such as the mean characteristic twist parameter MEANALP and the mean vertical current helicity MEANJZH;}
\item{For all SHARP parameters, the flaring HARPs tend to show higher $\alpha$ values than the flare-quiet HARPs;}
\item{For both the flare-quiet and flaring HARPs, most of the SHARP parameter time series show long-term (a few tens of hours to a few days) variations (\textit{i.e.} non-stationary);}
\item{Time series of the total unsigned magnetic flux USFLUX and the mean photospheric magnetic free energy density MEANPOT present a persistent property, while the total unsigned flux near magnetic polarity inversion lines R\_VALUE, the mean vertical current density MEANJZD, and the sum of the absolute value of the net current per polarity SAVNCPP show an anti-persistent trend in their time series.}
\end{enumerate}

Characteristic patterns of variation have been identified in time series of a variety of AR magnetic parameters over the dynamic evolution of flaring ARs \citep{2001SoPh..204...11D,2007A&A...474..633R,2008ApJ...686.1397P,2011ApJ...740...19R,2012ApJ...748...77S,2013RAA....13..226S,2014SoPh..289.2459S,2015RAA....15.1547V}. In this study, we found that most of the 8-day time series of the 18 SHARP parameters show long-term variations for both the flare-quiet and flaring HARPs. As reported, there is a 24-hour variation in the HMI magnetic field products due to the orbital velocity of SDO relative to the Sun \citep{2014SoPh..289.3483H,2016SoPh..291.1887C}. To understand how significant the 24-hour variation is and whether it can make a large influence in estimating $\alpha$ from the DFA method, we first determine the maximum absolute amplitude of spectral components with periods between 24$\pm$3 hours applying the Fast Fourier Transform (FFT) to the 8-day SHARP parameter time series. For all SHARP parameter time series, the median of the maximum absolute amplitude values is found to be $\sim$11\% of their interquartile ranges (\textit{i.e.} difference between the 1st and 3rd quartiles: $\mathrm{X}_\mathrm{3rd}$\,$-$\,$\mathrm{X}_\mathrm{1st}$). Then, for the set of the fBm and fGn time series used in Section~\ref{s:data}, we add a sinusoidal function with a period of 24 hours and an amplitude of 0.11\,$\times$\,$[\mathrm{X}_\mathrm{3rd}$\,$-$\,$\mathrm{X}_\mathrm{1st}]$. In the case of the fBm and fGn time series with 12 missing data points, $\alpha$ values calculated from the fBm and fGn added with the sinusoidal function are found to be slightly overestimated (0.01--0.09) for $H$\,$<$\,0.75 compared to the fBm and fGn without adding the sinusoidal function, while somewhat overestimated (0.09--0.27) for $H$\,$>$\,0.75. Note that $\alpha$ values are in general distributed in the range of $H$\,$<$\,0.75 for most SHARP parameters which show high levels of discriminating the flaring HARPs from the flare-quiet ones. It is therefore thought that the 24-hour variation which might be embedded in the SHARP parameter time series as a consequence of the SDO geosynchronous orbit does not significantly affect the main results in this study. In addition, we test how sensitive the estimation of $\alpha$ is to the centre-to-limb variation of the HMI observables, which is another systematic error embedded in $\textbf{X}$ \citep{2014SoPh..289.3483H}. For this, three 4-day segments are extracted from each of the 8-day SHARP parameter time series as follows: $\left[\mathrm{X}_{1},\mathrm{X}_{2},\mathrm{X}_{3},\ldots, \mathrm{X}_{480}\right]$, $\left[\mathrm{X}_{241},\mathrm{X}_{242},\mathrm{X}_{243},\ldots, \mathrm{X}_{720}\right]$, and $\left[\mathrm{X}_{481},\mathrm{X}_{482},\mathrm{X}_{483},\ldots, \mathrm{X}_{960}\right]$, corresponding to the intervals when the centre position of the given HARP is located within about $-60$--$0^\circ$, $-30$--$30^\circ$, and $0$--$60^\circ$, respectively. Estimating $\alpha$ for each segment (\textit{i.e.}, $\alpha_\mathrm{1st}$, $\alpha_\mathrm{2nd}$, and $\alpha_\mathrm{3rd}$ for the first, second, and third segments), we find that $\alpha_\mathrm{1st}-\alpha_\mathrm{2nd}$ and $\alpha_\mathrm{3rd}-\alpha_\mathrm{2nd}$ is, on average, 0.02 and 0.01, respectively, for all SHARP parameter time series studied here. The difference between $\alpha$ values estimated from the different segments is very small compared to $\alpha$ values themselves so that the $\alpha$ estimation is not thought to be significantly affected by the centre-to-limb variation.

For all SHARP parameters, the distribution of $\alpha$ shows based on the statistical significance test (\textit{i.e.} Student's $t$-test) that the flaring HARPs have relatively higher $\alpha$ values than the flare-quiet HARPs. This indicates that the temporal behavior of the SHARP parameter time series is different between the flare-quiet and the flaring HARPs. In other words, the flare-quiet HARPs tend to show larger fluctuations on shorter timescales (several tens of minutes to a few hours) and/or less fluctuations on longer timescales (24 hours to a few days), compared to the flaring HARPs. This may be again associated with the characteristic variations patterns (in general, long-term scales of a few days) which were found mainly in flaring ARs but not in flare-quiet ARs \citep[\textit{e.g.}][]{2008ApJ...686.1397P,2010ApJ...718...43P}. 

We find that the time series of some space-integrated, unsigned SHARP parameters, such as USFLUX, TOTPOT, and TOTUSJZ, show the persistent property, in particular for the flaring HARPs; these \textit{extensive} parameters reflect the size of ARs \citep{2009ApJ...705..821W}. On the other hand, space-averaged, signed parameters, such as MEANJZD, MEANJZH, and MEANALP, show the anti-persistent trend for the flare-quiet HARPs. In the case of flare-quiet ARs, it is typically shown that the spatial distribution of signed magnetic field properties is well balanced between positive and negative values \citep[\textit{e.g.}][]{2017SoPh..292..159K,2018SoPh..293...96K,2019SoPh..294..130K}, which may account for the anti-persistent property in their time series; \textit{i.e.} short-term fluctuation becomes larger in order to manage to balance the given signed property of positive and negative values. There have been several studies \citep[\textit{e.g.}][]{2015ApJ...798..135B,2019JKAS...52..133L} in which instantaneous and/or time-averaged values of space-averaged, signed photospheric magnetic field parameters are used to differentiate between the flare-quiet and flaring ARs; consequently, such parameters are found to be not as useful as space-integrated, unsigned parameters. However, in our study, space-averaged, signed SHARP parameters, such as MEANALP and MEANJZH, show statistically significant differences in $\alpha$ distributions between the flare-quiet and flaring HARPs, which suggests using scaling properties of fluctuations embedded in time series of AR magnetic field parameters may lead to improvement in performance of flare forecasting.

Finally, it will be interesting to examine whether/how scaling properties of time series fluctuations are related to spatial fractal properties in the context of AR flaring activity. For example, \citet{2012SoPh..276..161G} reported that there is no significant difference in fractal properties of photospheric line-of-sight magnetograms between flare-quiet and flaring ARs. In the time series of USFLUX under our study, we find a similar result that $\alpha$ values show little discrimination between the flare-quiet and flaring HARPs. Would spatial fractal properties be different between the flare-quiet and flaring HARPs in the case of two-dimensional maps for other SHARP parameters that explicitly show good discrimination? This question will be answered in our future study investigating, for example, current density or magnetic twist maps of active regions at various levels of flare activity.

\begin{acks}
The authors would like to thank an anonymous referee for valuable comments and suggestions. This work was supported by the BK21 plus program through the National Research Foundation (NRF) funded by the Ministry of Education of Korea, the Basic Science Research Program through the NRF funded by the Ministry of Education (NRF-2019R1A2C1002634), the Korea Astronomy and Space Science Institute (KASI) under the R\&D program "Study on the Determination of Coronal Physical Quantities using Solar Multi-wavelength Images" (project No. 2019-1-850-02) supervised by the Ministry of Science and ICT, and Institute for Information \& communications Technology Promotion (IITP) grant funded by the Korea government (MSIP) (2018-0-01422, Study on analysis and prediction technique of solar flares). The data used in this work are courtesy of the NASA/SDO and HMI science team. This research has made use of NASA’s Astrophysics Data System (ADS). S.-H.P. acknowledges support from the Institute for Space-Earth Environmental Research (ISEE) of Nagoya University as well as MEXT/JSPS KAKENHI Grant Number JP15H05814, Project for Solar-Terrestrial Environment Prediction (PSTEP).
\end{acks}
\hfill \break
\hfill \break
\noindent \textbf{Disclosure of Potential Conflicts of Interest} The authors declare that they have no conflicts of interest.

\bibliographystyle{spr-mp-sola.bst} 
\bibliography{ref.bib}

\clearpage

\begin{table}
\begin{adjustbox}{width=\textwidth, center=\textwidth}
\caption{Space-weather HMI Active Region Patch (SHARP) parameters}
\label{table_1}
\begin{tabular}{clll}
\hline
Keyword   & Description                                        & Unit    & Formula  \\ 
\hline
USFLUX    & Total unsigned magnetic flux                                & Mx       & $\Phi = \sum \left| B_z \right| dA$  \\
MEANGAM   & Mean angle of field from radial                    & Degrees  & $\overline{\gamma} = \frac{1}{N} \sum \arctan \left( \frac{B_h}{B_z} \right)$  \\
MEANGBT   & Mean gradient of total field                       & G Mm$^{-1}$   & $\overline{\left| \nabla B_{tot} \right|} = \frac{1}{N} \sum \sqrt{\left( \frac{\partial B}{\partial x} \right)^2 + \left( \frac{\partial B}{\partial y} \right)^2}$  \\
MEANGBZ   & Mean gradient of vertical field                    & G Mm$^{-1}$   & $\overline{\left| \nabla B_z \right|} = \frac{1}{N} \sum \sqrt{\left( \frac{\partial B_z}{\partial x} \right)^2 + \left( \frac{\partial B_z}{\partial y} \right)^2}$  \\
MEANGBH   & Mean gradient of horizontal field                  & G Mm$^{-1}$   & $\overline{\left| \nabla B_h \right|} = \frac{1}{N} \sum \sqrt{\left( \frac{\partial B_h}{\partial x} \right)^2 + \left( \frac{\partial B_h}{\partial y} \right)^2}$  \\
MEANJZD   & Mean vertical current density                      & mA m$^{-2}$   & $\overline{J_z} \propto \frac{1}{N} \sum \left( \frac{\partial B_y}{\partial x} - \frac{\partial B_x}{\partial y} \right)$  \\
TOTUSJZ   & Total unsigned vertical current                    & A        & $J_{z_{total}} = \sum \left| J_z \right| dA$  \\
MEANALP   & Mean characteristic twist parameter, $\alpha$      & Mm$^{-1}$     & $\alpha_{total} \propto \frac{\sum J_z \cdot B_z}{\sum B_z^2}$  \\
MEANJZH   & Mean vertical current helicity                     & G$^2$ m$^{-1}$   & $\overline{H_c} \propto \frac{1}{N} \sum B_z \cdot J_z$  \\
TOTUSJH   & Total unsigned vertical current helicity           & G$^2$ m$^{-1}$   & $H_{c_{total}} \propto \sum \left| B_z \cdot J_z \right|$  \\
ABSNJZH   & Absolute value of the net vertical current helicity & G$^2$ m$^{-1}$   & $H_{c_{abs}} \propto \left| \sum B_z \cdot J_z \right|$  \\
SAVNCPP   & Sum of the absolute value of the net current per polarity & A        & $J_{z_{sum}} \propto \left| \sum^{B_z^+} J_z dA \right| + \left| \sum^{B_z^-} J_z dA \right|$  \\
MEANPOT   & Mean photoshperic magnetic free energy density     & erg cm$^{-3}$ & $\overline{\rho} \propto \frac{1}{N} \sum \left( \mathbf{B}^{Obs} - \mathbf{B}^{Pot} \right)^2$  \\
TOTPOT    & Surface integral of photospheric magnetic free energy density    & erg cm$^{-1}$ & $\rho_{tot} \propto \sum \left( \mathbf{B}^{Obs} - \mathbf{B}^{Pot} \right)^2 dA$  \\
MEANSHR   & Mean shear angle                                   & Degrees  & $\overline{\Gamma} = \frac{1}{N} \sum \arccos \left( \frac{\mathbf{B}^{Obs} \cdot \mathbf{B}^{Pot}}{\left| B^{Obs} \right| \left| B^{Pot} \right|} \right)$  \\
SHRGT45   & Fraction of Area with shear > 45$^\circ$       &          & Area with shear > 45$^\circ$ / total area  \\
R\_VALUE  & Sum of flux near polarity inversion line           & Mx       & $\Phi = \sum \left| B_{LoS} \right| dA$ within $R$ mask  \\
AREA\_ACR & Area of strong field pixels in the active region   &          & Area = $\sum \text{Pixels}$ \\
\hline
\end{tabular}
\end{adjustbox}
\begin{tablenotes}
\tiny
\item Note: Further description of the SHARP parameters can be found in \citet{2015ApJ...798..135B} and references therein.
\end{tablenotes}
\end{table}

\begin{table}
\begin{adjustbox}{width=\textwidth, center=\textwidth}
\caption{Percentage distribution of the scaling exponent $\alpha$}
\label{table_2}
\begin{tabular}{cccccccccc}
\hline
\multirow{3}{*}{Type} & \multirow{2}{*}{SHARP} & \multicolumn{4}{c}{Flare-quiet HARPs} & \multicolumn{4}{c}{Flaring HARPs} \\ \cmidrule(rl){3-6} \cmidrule(rl){7-10}
 & Parameter & \multirow{2}{*}{Stationary} & \multicolumn{3}{c}{Non-stationary} & \multirow{2}{*}{Stationary} & \multicolumn{3}{c}{Non-stationary} \\ \cmidrule(r){4-6} \cmidrule(r){8-10} 
 & & & Anti-persistent & Random & Persistent & & Anti-persistent & Random & Persistent \\ \hline
\multirow{2}{*}{\textit{PP}}   & USFLUX       & 0       & 0.05    & 0.14    & \textit{\textbf{0.81}}    & 0       & 0.06    & 0.07    & \textit{\textbf{0.87}} \\
                               & MEANPOT      & 0.01    & 0.04    & 0.30    & \textit{\textbf{0.65}}    & 0.02    & 0.06    & 0.17    & \textit{\textbf{0.75}} \\ \hline
\multirow{5}{*}{\textit{RP}}   & MEANGBH      & 0       & 0.06    & \textit{\textbf{0.49}}    & 0.45    & 0.01    & 0.03    & 0.12    & \textit{\textbf{0.84}} \\
                               & MEANGBT      & 0.02    & 0.16    & \textit{\textbf{0.64}}    & 0.19    & 0.02    & 0.08    & 0.25    & \textit{\textbf{0.66}} \\
                               & MEANGBZ      & 0.03    & 0.17    & \textit{\textbf{0.59}}    & 0.20    & 0.02    & 0.10    & 0.19    & \textit{\textbf{0.69}} \\
                               & TOTPOT       & 0.02    & 0.10    & \textit{\textbf{0.55}}    & 0.33    & 0       & 0.05    & 0.38    & \textit{\textbf{0.57}} \\
                               & TOTUSJZ      & 0.08    & 0.25    & \textit{\textbf{0.52}}    & 0.15    & 0.08    & 0.15    & 0.35    & \textit{\textbf{0.42}} \\ \hline
\multirow{3}{*}{\textit{RR}}   & TOTUSJH      & 0       & 0.19    & \textit{\textbf{0.66}}    & 0.14    & 0       & 0.09    & \textit{\textbf{0.49}}    & 0.42 \\
                               & AREA\_ACR    & 0       & 0.04    & \textit{\textbf{0.89}}    & 0.07    & 0       & 0       & \textit{\textbf{0.71}}    & 0.29 \\
                               & MEANSHR      & 0.10    & 0.43    & \textit{\textbf{0.47}}    & 0       & 0.04    & 0.27    & \textit{\textbf{0.65}}    & 0.03 \\ \hline
\multirow{5}{*}{\textit{AR}}   & MEANGAM      & 0.04    & \textit{\textbf{0.48}}    & 0.47    & 0       & 0       & 0.20    & \textit{\textbf{0.76}}    & 0.03 \\
                               & SHRGT45      & 0.09    & \textit{\textbf{0.51}}    & 0.40    & 0       & 0.03    & 0.35    & \textit{\textbf{0.58}}    & 0.04 \\
                               & MEANJZH      & 0.01    & \textit{\textbf{0.74}}    & 0.25    & 0       & 0       & 0.18    & \textit{\textbf{0.70}}    & 0.12 \\
                               & ABSNJZH      & 0.01    & \textit{\textbf{0.78}}    & 0.21    & 0       & 0       & 0.22    & \textit{\textbf{0.75}}    & 0.03 \\
                               & MEANALP      & 0.02    & \textit{\textbf{0.90}}    & 0.08    & 0       & 0       & 0.25    & \textit{\textbf{0.73}}    & 0.02 \\ \hline
\multirow{3}{*}{\textit{AA}}   & SAVNCPP      & 0.18    & \textit{\textbf{0.80}}    & 0.03    & 0       & 0       & \textit{\textbf{0.85}}    & 0.15    & 0    \\
                               & R\_VALUE     & 0.14    & \textit{\textbf{0.85}}    & 0.02    & 0       & 0       & \textit{\textbf{0.89}}    & 0.10    & 0.01 \\
                               & MEANJZD      & 0.11    & \textit{\textbf{0.89}}    & 0       & 0       & 0.04    & \textit{\textbf{0.87}}    & 0.08    & 0    \\ \hline
\end{tabular}
\end{adjustbox}
\begin{tablenotes}
\tiny
\item Note: The four subgroups are defined as ranges of their $\alpha$ values as follows: stationary (0.6\,$\le$\,$\alpha$\,$<$\,1.0), non-stationary and anti-persistent (1.0\,$\le$\,$\alpha$\,$<$\,1.4), non-stationary and random (1.4\,$\le$\,$\alpha$\,$<$\,1.6), and non-stationary and persistent (1.6\,$\le$\,$\alpha$\,$<$\,2.0). The type (\textit{first column}) of each SHARP parameter is defined, considering the subgroup with the highest percentage (marked in bold) for the flare-quiet (first letter) and flaring HARPs (second letter), respectively.
\end{tablenotes}
\end{table}

\begin{table}
\begin{adjustbox}{width=7cm}
\caption{Student's $t$-test results}
\label{table_3}
\begin{tabular}{lcccc}
\hline
\multirow{2}{*}{SHARP Parameter} & \multicolumn{2}{c}{$t$-value} & \multicolumn{2}{c}{$p$-value$^\ddagger$} \\ \cmidrule(rl){2-3} \cmidrule(rl){4-5} 
 & $\alpha$    & $\mathrm{\bar{X}}$    & $\alpha$    & $\mathrm{\bar{X}}$    \\ \hline
\textbf{USFLUX}$^\dagger$       & 3.13         & \textbf{13.12}        & $<$\,0.01        &  $<$\,0.01       \\
MEANGAM      & 6.08         & 8.44         &  $<$\,0.01       &  $<$\,0.01       \\
MEANGBT      & 6.45         & -6.88        &  $<$\,0.01       &  $<$\,0.01       \\
MEANGBZ      & 6.58         & -5.13        &  $<$\,0.01       &  $<$\,0.01       \\
MEANGBH      & 6.63         & 4.39         &  $<$\,0.01       &  $<$\,0.01       \\
MEANJZD      & 6.60         & -6.82        &  $<$\,0.01       &  $<$\,0.01       \\
\textbf{TOTUSJZ}$^\dagger$      & 1.84         & \textbf{14.06}        & 0.07        &   $<$\,0.01      \\
\textbf{MEANALP}$^\ast$      & \textbf{14.79}        & 3.72         &  $<$\,0.01       &    $<$\,0.01     \\
\textbf{MEANJZH}$^\ast$      & \textbf{12.43}        & 4.74         &   $<$\,0.01      &  $<$\,0.01       \\
\textbf{TOTUSJH}$^\dagger$      & 5.31         & \textbf{13.27}        &  $<$\,0.01       &   $<$\,0.01      \\
\textbf{ABSNJZH}$^\ast$      & \textbf{12.11}        & 9.63         &  $<$\,0.01       &   $<$\,0.01      \\
\textbf{SAVNCPP}$^\ast$      & \textbf{14.12}       & 10.61        &     $<$\,0.01    &    $<$\,0.01     \\
MEANPOT      & 2.41         & 6.83         & 0.02        &   $<$\,0.01      \\
TOTPOT       & 4.64         & 9.47         &  $<$\,0.01       &   $<$\,0.01     \\
MEANSHR      & 3.96         & 9.21         &  $<$\,0.01       &  $<$\,0.01       \\
SHRGT45      & 5.33         & 8.52         &  $<$\,0.01       &  $<$\,0.01       \\
R\_VALUE     & 11.17        & 10.05        &  $<$\,0.01       &  $<$\,0.01       \\
\textbf{AREA\_ACR}$^\dagger$    & 7.01         & \textbf{13.21}        &  $<$\,0.01       &   $<$\,0.01      \\ \hline
\end{tabular}
\end{adjustbox}
\begin{tablenotes}
\tiny
\item $^\ast$ Four best-discriminating parameters for $\alpha$
\item $^\dagger$ Four best-discriminating parameters fo $\mathrm{\bar{X}}$
\item $^\ddagger$ $p$-values less than 0.01 are marked as $<$\,0.01. 
\end{tablenotes}
\end{table}

\end{article} 
\end{document}